\documentclass[12pt,a4paper]{article}
\usepackage{geometry}
\usepackage{jheppub} 
\usepackage{comment}
\usepackage[english]{babel}
\usepackage[labelfont=bf]{caption}
\usepackage{empheq}
\usepackage{bbm}
\usepackage{float}
\usepackage{comment}
\usepackage[utf8]{inputenc}
\usepackage{natbib} 
\usepackage{mathrsfs}
\usepackage[mathscr]{eucal}
\usepackage{amsmath}
\usepackage{amsfonts}
\usepackage[most]{tcolorbox}
\usepackage{amssymb}
\usepackage{graphicx}
\usepackage{scalerel}
\usepackage{multirow}
\usepackage{cancel}
\usepackage{slashed}
\usepackage{booktabs}
\usepackage{braket}
\usepackage{hyperref}
\usepackage{bm}
\usepackage{multicol}
\usepackage{multirow}
\usepackage{wasysym}

\DeclareCaptionLabelFormat{andtable}{#1~#2  \&  \tablename~\thetable} 

\allowdisplaybreaks 
\usepackage{fancyhdr}
\usepackage{graphicx,scalerel}

\usepackage[shortlabels]{enumitem}
\setlist{itemsep=.1em,topsep=.5em}
\SetEnumerateShortLabel{i}{\textit{\roman*}}

\numberwithin{equation}{section}
\numberwithin{figure}{section}
\numberwithin{table}{section}

\usepackage[noindentafter]{titlesec} 
\DeclareMathAlphabet{\mathsfit}{OT1}{lmss}{m}{sl}
\DeclareMathAlphabet{\mathsfbf}{OT1}{lmss}{bx}{n}
\DeclareMathAlphabet{\mathsfbfit}{OT1}{lmss}{bx}{sl}
\DeclareMathVersion{chaptermath}
\SetSymbolFont{operators}{chaptermath}{OT1}{cmbr}{m}{n}
\SetSymbolFont{letters}{chaptermath}{OML}{cmbrm}{b}{it}
\SetSymbolFont{symbols}{chaptermath}{OMS}{cmbrs}{m}{n}
\DeclareMathVersion{subsectionmath}
\SetSymbolFont{operators}{subsectionmath}{OT1}{cmbr}{m}{n}
\SetSymbolFont{letters}{subsectionmath}{OML}{cmbrm}{m}{it}
\SetSymbolFont{symbols}{subsectionmath}{OMS}{cmbrs}{m}{n}

\titleformat{\chapter}{\bfseries \huge  }{\thechapter. }{-1pt}{}{}
\titlespacing{\chapter}{0pt}{10pt plus 1pt minus 1pt}{10pt plus 1pt minus 1pt}

\titleformat{\section}{\large \bfseries   }{\thesection}{10pt}{}{}
\titlespacing{\section}{0pt}{15pt plus 1pt minus 1pt}{5pt plus 1pt minus 1pt}
\titleformat{\subsection}{\normalsize \bfseries  }{\thesubsection}{10pt}{}{}
\titlespacing{\subsection}{0pt}{15pt plus 1pt minus 1pt}{5pt plus 1pt minus 1pt}
\titleformat{\subsubsection}{\normalsize \bfseries    }{\thesubsubsection}{10pt}{}{}
\titlespacing{\subsubsection}{0pt}{15pt}{5pt}

\begin{document}

\renewcommand{\headrulewidth}{0pt}
\renewcommand{\footrulewidth}{0pt}

\begin{flushright}
MITP-25-041\\
June 6, 2025
\end{flushright}

\thispagestyle{empty}
\begin{center}
\vspace{0.8cm}
{\Large\bf Saving or Destroying the Universe \\[3mm]with Axion-Like Particles
}

\vspace{1cm}
{Anne Mareike Galda$^a$ and Matthias Neubert$^{a,b}$}\\[0.2cm]
{\em ${}^a$\,PRISMA$^+$ Cluster of Excellence {\em \&} MITP, 
Johannes Gutenberg University\\ 
55099 Mainz, Germany\\[2mm]
${}^b$\,Department of Physics {\em \&} LEPP, Cornell University, Ithaca, NY 14853, U.S.A.}
\end{center}
\vspace{3mm}

\begin{center}
\textbf{Abstract}
\end{center}
\vspace{-0.2cm}

\begin{abstract}
\,\noindent Light pseudoscalar resonances that couple to the Standard Model via non-ren\-ormaliz\-able operators, such as axions and axion-like particles (ALPs), generate contributions to the renormalization group evolution equations of couplings of dimension-4 and higher-dimensional operators. In particular, they modify the $\beta$-function of the Higgs quartic coupling and of SM and SMEFT parameters entering this equation, thus having an impact on the instability scale of the electroweak vacuum. We employ this fact together with the requirement that, in the presence of axions and ALPs, the Universe remains in a meta-stable state to deduce bounds on ALP couplings to the Standard Model fields. We also show that the modification of the $\beta$-functions of the gauge couplings by the ALP can lead to a unification around the Planck scale, even in non-sypersymmetric models.
\end{abstract}

\section{Introduction}
Axions and axion-like particles (ALPs) are theoretically well-motivated candidates for physics extending the particle content of the Standard Model (SM). They arise as pseudo Nambu-Goldstone bosons from the breaking of an additional global $U(1)_A$ symmetry. As such, they can provide elegant answers to some fundamental questions regarding the SM, such as the strong CP problem \cite{Peccei:1977hh, Weinberg:1977ma, Wilczek:1977pj}, the flavor puzzle \cite{Calibbi:2016hwq,Ema:2016ops} and the hierarchy problem \cite{Bagger:1994hh, Gripaios:2009pe, Ferretti:2013kya, Graham:2015cka, Bellazzini:2017neg}. Due to their shift-symmetric interactions with the SM particles, operators that describe these couplings start at dimension-5 order. This leads to the phenomenon of the ALP--SMEFT interference described in \cite{Galda:2021hbr}: due to one-loop virtual ALP-exchange, nonzero dimension-6 SMEFT Wilson coefficients are generated at the scale of global symmetry breaking independently of the ALP mass. In addition, the presence of the ALP also modifies the renormalization-group (RG) evolution equations of the renormalizable couplings. One parameter that receives such an ALP-generated term in its RG equation is the quartic coupling $\lambda$ of the Higgs doublet. While in the SM the measured value of the Higgs mass leads to a scenario in which the Universe is in a meta-stable state, in which the lifetime of the electroweak vacuum exceeds the age of the Universe \cite{Degrassi:2012ry}, any modification of the $\beta$-function of $\lambda$ from physics beyond the SM can potentially either relax or tighten this scenario. Requiring that the presence of ALPs does not destabilize the Universe, it is thus possible to deduce bounds on those couplings to the SM particles that have a nontrivial effect on the evolution of $\lambda$, which in this paper we do for a generic (i.e.~model-independent) ALP. In Section \ref{sec:ALP_Lagrangian} we introduce the general ALP--SM Lagrangian up to dimension-6 order, followed by a brief summary of the concept of the ALP--SMEFT interference in Section~\ref{sec:ALP_SMEFT_Interference}. Subsequently, we review the estimate of the Universe's lifetime based on the tunneling probability to the true ground state in Section~\ref{sec:Instability_Scale}. In Section~\ref{sec:Instability_Scale_Mod}, the effects of single ALP couplings on the Higgs quartic coupling are analyzed, and the constraints that can be obtained from this consideration are presented in  Section~\ref{sec:Instability_Regions}. In a short outlook in Section~\ref{sec:GUT} we furthermore show that the presence of ALPs in the non-supersymmetric SM modifies the $\beta$-functions of the three gauge couplings $\alpha_1,\,\alpha_2$ and $\alpha_3$ such that a unified value at around $10^{15}$~GeV can be achieved. 
While we remain agnostic about the UV physics, employing concrete models for the ALP, such as for instance the KSVZ or DFSZ models, would modify our results in two ways: First, threshold corrections at the $U(1)_A$ symmetry breaking scale could impact $\lambda$. Second, the new degrees of freedom might contribute additional terms to the running of the Higgs quartic coupling. Thus, the results of this work are not ultimate bounds on ALP couplings, but rather show the general concept of ALP solutions to fundamental questions posed by the SM.

\section{ALP couplings to the SM particles}
\label{sec:ALP_Lagrangian}
We consider a gauge-singlet, pseudoscalar ALP whose classical shift symmetry allows for operators with couplings to the SM fields starting at dimension-5 order. The suppression scale is given by the scale of global (Peccei--Quinn) symmetry breaking $\Lambda = 4 \pi f \gg m_a$, where $f$ denotes the ``ALP decay constant" and we allow for a soft breaking of the shift symmetry by an explicit mass term $m_a$. 
Concretely, the most general Lagrangian for such a resonance $a$ up to dimension-6 order is given by \cite{Georgi:1986df}
\begin{align}
\label{eq:lag1}
   \mathcal{L}_\mathrm{SM+ALP}^{D\le 6}=&  \,c_{GG}\,\frac{a}{f}\,\frac{\alpha_s}{4\pi}\,G_{\mu\nu}^A\,\tilde G^{\mu\nu\,A} + c_{WW}\,\frac{\alpha_L}{4\pi}\,\frac{a}{f}\,W_{\mu\nu}^I\,\tilde W^{\mu\nu\,I} + c_{BB}\,\frac{\alpha_Y}{4\pi}\,\frac{a}{f}\,B_{\mu\nu}\,\tilde B^{\mu\nu} \nonumber\\
   & + \frac{\partial^\mu a}{f}\,\sum_F \bar \psi_F\, \bm{c}_F \,\gamma_\mu\, \psi_F + \frac{C_{HH}}{f^2}\,(\partial^\mu a)(\partial_\mu a)\,H^\dagger H\,.
\end{align}
In this expression, $G_{\mu\nu}^A,\,W_{\mu\nu}^I$ and $B_{\mu\nu}$ are the field-strength tensors of $SU(3)_c$, $SU(2)_L$ and $U(1)_Y$, respectively, and $\tilde G^{\mu\nu,\,A} = \frac{1}{2} \epsilon^{\mu\nu\alpha\beta}G_{\alpha\beta}^A$ etc.~denote their duals. The couplings to the SM fermions $F = Q,\,L,\,u,\,d,\,e$ are described in terms of $3\times 3$ hermitian matrices $\bm{c}_F$ in generation space. In order to remove the derivative interaction of the ALP to the fermions, we perform a field redefinition of the form $\psi_f\to \psi_f + i\,\frac{a}{f}\,\bm{c}_f\,\psi_f$, yielding the effective Lagrangian \cite{Bauer:2020jbp}
\begin{align}\label{eq:lag2}
\mathcal{L}_\mathrm{SM+ALP}^{D\le 6}&\to C_{GG}\,\frac{a}{f}\,G_{\mu\nu}^A\,\tilde G^{\mu\nu\,A} + C_{WW}\,\frac{a}{f}\,W_{\mu\nu}^I\,\tilde W^{\mu\nu\,I} + C_{BB}\,\frac{a}{f}\,B_{\mu\nu}\,\tilde B^{\mu\nu} \nonumber\\
&\quad - \frac{a}{f} \left( \bar{Q} \, \tilde{H} \,\tilde{\bm{Y}}_u \, u_R + \bar{Q} \, H \,\tilde{\bm{Y}}_d \, d_R + \bar{L} \, H \,\tilde{\bm{Y}}_e \, e_R + \mathrm{h.c.}\right) \nonumber\\
&\quad + \frac{1}{2}\,\frac{a^2}{f^2}\,\left(\bar Q\,\tilde{H}\,\bm{Y}_u^\prime \,u_R +\bar Q\,H\,\bm{Y}_d^\prime \,d_R +\bar L\, H\, \bm{Y}_e^\prime \,e_R + \mathrm{h.c.}\right)\nonumber\\
&\quad+ \frac{C_{HH}}{f^2}\,(\partial^\mu a)(\partial_\mu a)\,H^\dagger H\,,
\end{align}
with 
\begin{align}
\label{eq:Yukawa_tile_prime}
\tilde{\bm{Y}}_{F_R} &= i\, (\bm{Y}_{F_R} \,\bm{c}_{F_R} - \bm{c}_{F_L}\, \bm{Y}_{F_R} )\,, \notag\\
\bm{Y}_{F_R}^\prime &= \bm{c}_{F_L}^2\, \bm{Y}_{F_R} - 2\, \bm{c}_{F_L}\,\bm{Y}_{F_R}\,\bm{c}_{F_R} + \bm{Y}_{F_R}\,\bm{c}_{F_R}^2\,,
\end{align} 
where $\bm{Y}$ denotes a SM Yukawa matrix with $F_R = u,\,d,\,e$ and $F_L = Q,\,L$. The ALP--boson couplings in \eqref{eq:lag1} are related to those in \eqref{eq:lag2} by
\begin{align}
\label{eq:related_coefs_bosons}
    C_{GG} &= \frac{\alpha_s}{4\pi}\,\left[ c_{GG} + \frac{1}{2}\,\text{Tr}(\bm{c}_d  +\bm{c}_u - 2\bm{c}_Q) \right] , \notag\\
    C_{WW} &= \frac{\alpha_L}{4\pi}\,\left[ c_{WW} - \frac{1}{2}\,\text{Tr}(N_c\,\bm{c}_Q  +\bm{c}_L) \right] ,\nonumber\\
    C_{BB} &= \frac{\alpha_Y}{4\pi}\,\left[ c_{BB} +\text{Tr}\Big(N_c \,(\mathcal{Y}_d^2\,\bm{c}_d + \mathcal{Y}_u^2\,\bm{c}_u - 2 \,\mathcal{Y}_Q^2\,\bm{c}_Q)  + \mathcal{Y}_e^2\,\bm{c}_e - 2\,\mathcal{Y}_L^2\,\bm{c}_L \Big) \right] \,,
\end{align}
where $\mathcal{Y}_u = \frac{2}{3}$, $\mathcal{Y}_d = -\frac{1}{3}$, $\mathcal{Y}_Q = \frac{1}{6}$, $\mathcal{Y}_e = -1$ and $\mathcal{Y}_L = -\frac{1}{2}$ denote the hypercharge quantum numbers of the SM quarks and leptons.
In the following, we will assume a flavor--universal scenario for the ALP-fermion interactions. Thus, we set $\bm{c}_F = c_F\mathbbm{1}_3$ and define $C_{u,\,d} \equiv c_{u,\,d} - c_{q} $ and $C_{e} \equiv c_{e} - c_{l} $.
 
\section{ALP--SMEFT Interference}
\label{sec:ALP_SMEFT_Interference}
As the ALP-SM interactions start at dimension-5 order, one-loop graphs with virtual ALP exchange and external SM particles generate UV divergences that require dimension-6 counterterms built out of SM fields. Even if the ALP is very light and thus not integrated out, additional, inhomogeneous source terms, generated at the scale of the global $U(1)_A$ symmetry breaking, need to be added to the RG evolution equations of the SMEFT Wilson coefficients. In detail, in the presence of an ALP, the scale evolution of the SMEFT coefficients can be written as \cite{Galda:2021hbr}
\begin{align}
\label{eq:lambda_contribution}
     \frac{d}{d\ln\mu} \,C^{\mathrm{SMEFT}}_i - \gamma_{ji}^{\mathrm{SMEFT}} C_j^{\mathrm{SMEFT}} = \frac{S_i}{(4\pi f)^2}\,,\quad \text{for}\quad\mu_w < \mu < 4\pi f\,,
\end{align}
where $\mu_w$ denotes the scale of electroweak symmetry breaking. 
The full list of the source terms $S_i$ has been obtained in \cite{Galda:2021hbr}.\\
In addition, one-loop amplitudes proportional to $m_a^2/f^2$ and $m_H^2/f^2$ modify the RG flow of Wilson coefficients of renormalizable operators. In particular, the quartic Higgs coupling $\lambda$ receives a contribution of the form \cite{Galda:2021hbr}
\begin{align}
\label{eq:CWW_in_lambda}
    \frac{d \lambda}{d\ln\mu} \supset -\frac{16 g_s^2}{3} \,\frac{m_H^2}{(4\pi f)^2}\, C_{WW}^2\,,
\end{align}
while the Higgs mass parameter gets a contribution from the dimension-6 operator in \eqref{eq:lag1} that reads \cite{Biekotter:2023mpd}
\begin{align}
      \frac{d m_H^2}{d\ln\mu} \supset - \frac{m_a^4}{(4\pi f)^2} \,C_{HH}\,.
\end{align}
\begin{figure}[t]
\centering
\includegraphics[width=.7\textwidth]{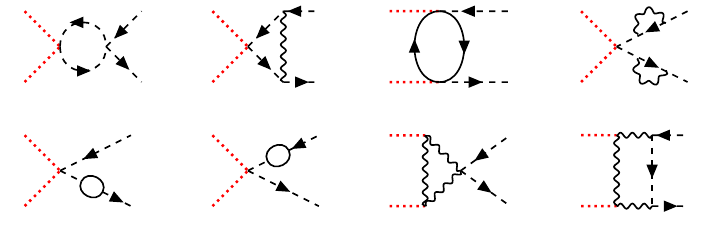}
\caption{Feynman diagrams contributing to the RG evolution of the coefficient $C_{HH}$ multiplying the dimension-6 operator that contains two ALP and two Higgs fields as defined in \eqref{eq:lag1}. A red dotted line represents the ALP, while the Higgs fields are shown as black dashed lines with the arrow pointing in the direction of the $SU(2)_Y$ charge flow.}
\label{fig:RGE_Diagrams}
\end{figure}
Despite being the coefficient of a dimension-6 operator, $C_{HH}$ can thus modify SM parameters with the same power suppression in $f$ as the remaining dimension-5 ALP coefficients, which enter through the generation of dimension-6 SMEFT coefficients. Thus, to fully solve the coupled set of equations, we also need the dependence of $C_{HH}$ on the RG scale $\mu$ which has not yet been derived in the literature. We find that the diagrams contributing to the running are those depicted in Figure~\ref{fig:RGE_Diagrams}, yielding, for $N_c = 3$, a $\beta$-function of the form
\begin{align}
    \frac{d \,C_{HH}}{d\ln \mu} = \frac{1}{(4 \pi)^2}&\bigg( 36\, g_L^2 \,C_{WW}^2  + 12\, g_Y^2 \, C_{BB}^2- 6\, \mathrm{Tr}[\tilde{\bm{Y}}_d \tilde{\bm{Y}}_d^\dagger + \tilde{\bm{Y}}_u \tilde{\bm{Y}}_u^\dagger  + \frac{1}{3}\,\tilde{\bm{Y}}_e \tilde{\bm{Y}}_e^\dagger] \notag\\
    & - \frac{3}{2}\, C_{HH}\,\big(   3\, g_L^2 + g_Y^2-4\,\lambda - 4 \, \mathrm{Tr}[\bm{Y}_d \bm{Y}_d^\dagger + \bm{Y}_u \bm{Y}_u^\dagger  + \frac{1}{3}\,\bm{Y}_e \bm{Y}_e^\dagger]\big) \bigg) \,.
\end{align}
In the RG equation of $\lambda$, the explicit Higgs mass parameter $m_H$ enters only via dimension-6 SMEFT coefficients \cite{Jenkins:2013wua}, or via the term in \eqref{eq:CWW_in_lambda}. Thus, it is at least two-loop power suppressed and does not have a significant effect on the scale evolution of the Higgs quartic coupling. We will therefore drop it in the following discussion. 

\section{Instability of the electroweak vacuum in the Standard Model and beyond}
\label{sec:Instability_Scale}
With the measured value of the Higgs mass of $125.25\, \mathrm{GeV}$ \cite{ParticleDataGroup:2022pth}, the SM is in a near-critical situation in which the electroweak vacuum does not correspond to the absolute minimum of the Higgs potential, 
\begin{align}
    V = \frac{\lambda}{2} \,(H^\dagger H)^2-m_H^2 \,H^\dagger H\,,
\end{align}
but to a metastable state whose lifetime, however, exceeds the age of the Universe. The instability scale $\Lambda_I$ corresponds to the scale where the value of the potential becomes smaller than the electroweak vacuum, making it a false minimum with nonzero probability of tunneling to the true ground state. At very large field values where the Higgs mass can be neglected and the field value of the Higgs is approximated by the RG scale $\mu$, this corresponds roughly to the scale where $\lambda$ becomes negative. Including one-loop RGEs for the parameters this happens in the SM around $10^{11}\,\mathrm{GeV}$.\\
The presence of the ALP modifies the RG evolution equations of the quartic coupling two-fold: first, a direct contribution proportional to the coupling of the ALP to the $W$-boson as generated at the symmetry breaking scale $\Lambda$ enters the equation for $\lambda$. As this additional term is negative for any nonzero value of $C_{WW}$, it will necessarily push the instability scale towards lower values. Secondly, due to one-loop virtual ALP exchange, which we refer to as the \emph{ALP--SMEFT interference}, RG evolution equations of the dimension-4 SM couplings are modified, and non-zero SMEFT Wilson coefficients are generated. Thus, it does not suffice to consider the pure SM running only, but the full SMEFT running, which has been obtained in \cite{Jenkins:2013zja, Jenkins:2013wua, Alonso:2013hga}, needs to be taken into account. \\
In detail, we proceed as follows: for a given scale of global symmetry breaking $\Lambda = 4 \pi \,f$, ALP mass $m_a$ and ALP--SM couplings at $\Lambda$, we obtain the SMEFT and ALP couplings at $M_Z$ by solving the full set of SMEFT RG evolution equations. Depending on the ALP mass, this either means running the full dimension-6 system including the ALP coefficients down to the scale where $\lambda$ is measured (for $m_a \leq M_Z$), or integrating out the ALP in an intermediate step (which we preform at tree level) and continue with the pure SMEFT running without ALP modifications. The thus obtained initial conditions of the dimension-5 ALP and dimension-6 SMEFT parameters are then used as inputs at the appropriate scale (i.e.~ALP couplings only contribute for $\mu \geq m_a$). \\
The concrete effect of the ALP generated SMEFT contributions is non-trivial and depends on the configuration of ALP--SM couplings at the scale of global symmetry breaking $\Lambda$. This is the case, as the RG flow generates different sets of SMEFT and ALP couplings at low scales, and modifications to the SM couplings can thus be generated even without the direct contribution of $C_{WW}$ in \eqref{eq:CWW_in_lambda}. While one would maybe expect that this direct ALP contribution dominates the evolution of $\lambda$ in the presence of the ALP and that one-loop suppressed ALP contributions play a minor role in the evolution of $\lambda$, terms that are proportional to $m_a^2$ can get strongly enhanced for large ALP masses as we will show later.\\
This setup puts us in a position where we can constrain the ALP-SM couplings from the requirement that the electroweak vacuum is supposed to remain in a meta-stable state in the presence of the ALP, i.e.~the resulting lifetime remains larger than the age of the Universe.
To estimate the lifetime of the electroweak vacuum $\tau_{\mathrm{EW}}$ we follow the results obtained in \cite{Buttazzo:2013uya}. For completeness, we summarize their findings relevant for the following analysis.\\
Approximating the Universe by a sphere with radius $c\, T_U$, where the age of the Universe $T_U$ and the Hubble constant $H_0$ are related by $T_U \approx 0.96\,H_0$, the probability for the vacuum to decay to the true ground state is given by
\begin{align}
    p_0 = 0.15 \,\frac{\Lambda_B^4}{H_0^4}\,e^{-S(\Lambda_B)}\,.
\end{align}
Here, the action $S$ is related to the quartic Higgs coupling via
\begin{align}
    S(\Lambda_B) = \frac{16 \pi^2}{3 |\lambda(\Lambda_B)|}\,,
\end{align}
and $\Lambda_B$ corresponds to the scale where $\lambda(\Lambda_B)$ is minimized. In a vacuum-energy dominated Universe, the lifetime of the electroweak vacuum then reads
\begin{align}
\label{eq:tau_ew}
    \tau_{\mathrm{EW}} = \frac{3 H^3 e^{S(\Lambda_B)}}{4 \pi \Lambda_B^4}\approx \frac{0.02\,T_U}{p_0}\,.
\end{align}
The tunneling probability to the true electroweak vacuum thus depends on both the scale where $\lambda$ is minimized, as well as on the magnitude of $\lambda$ at this point. 

\section{Effects of the ALP on the quartic Higgs coupling}
\label{sec:Instability_Scale_Mod}
In the pure SM scenario without new physics, the instability scale of the electroweak vacuum is approximately given by $\Lambda_I =  10^{11}\,\mathrm{GeV}$. We now proceed by analyzing the question how the different ALP couplings influence this scale. To this end, we first investigate the two cases $m_a = 0$ and $m_a = 100\,\mathrm{GeV}$ and turn on one coupling at a time at the scale of global $U(1)_A$ symmetry breaking $\Lambda$. Employing the RG evolution equations for the ALP coefficients \cite{Bauer:2020jbp} and SMEFT Wilson coefficients \cite{Jenkins:2013wua,Alonso:2013hga} in a modified version of the \texttt{DsixTools} framework and the input values provided therein \cite{Celis:2017hod, Fuentes-Martin:2020zaz}, we obtain the numerical values of all relevant coefficients for the scale evolution of $\lambda$ at the electroweak scale, keeping $\lambda(M_Z) \approx 0.28$ fixed. Clearly, the magnitudes of the ALP-generated SMEFT Wilson coefficients at the electroweak scale depend on the scale evolution from $\Lambda$ down to $\mu$. In other words, when the symmetry breaking happens at a very high scale, the ratio $\Lambda/\mu$ yields large logarithms multiplying the SMEFT and ALP coefficients, thus enhancing the effect.  
As an example, we here set $f = 1\,\mathrm{TeV}$ and show the results for the modified instability scale for various ALP couplings in Figure~\ref{fig:Instability_Scale}. As expected from the direct contribution of $ C_{WW}$ to the RG evolution of $\lambda$ in \eqref{eq:CWW_in_lambda}, turning on the couplings to $W$-bosons has the effect of lowering the instability scale $\Lambda_I$. While it is straightforward to understand this behavior, disentangling the effects of the remaining parameters is more subtle:
\begin{figure}[t]
\centering
\includegraphics[width=1.\textwidth]{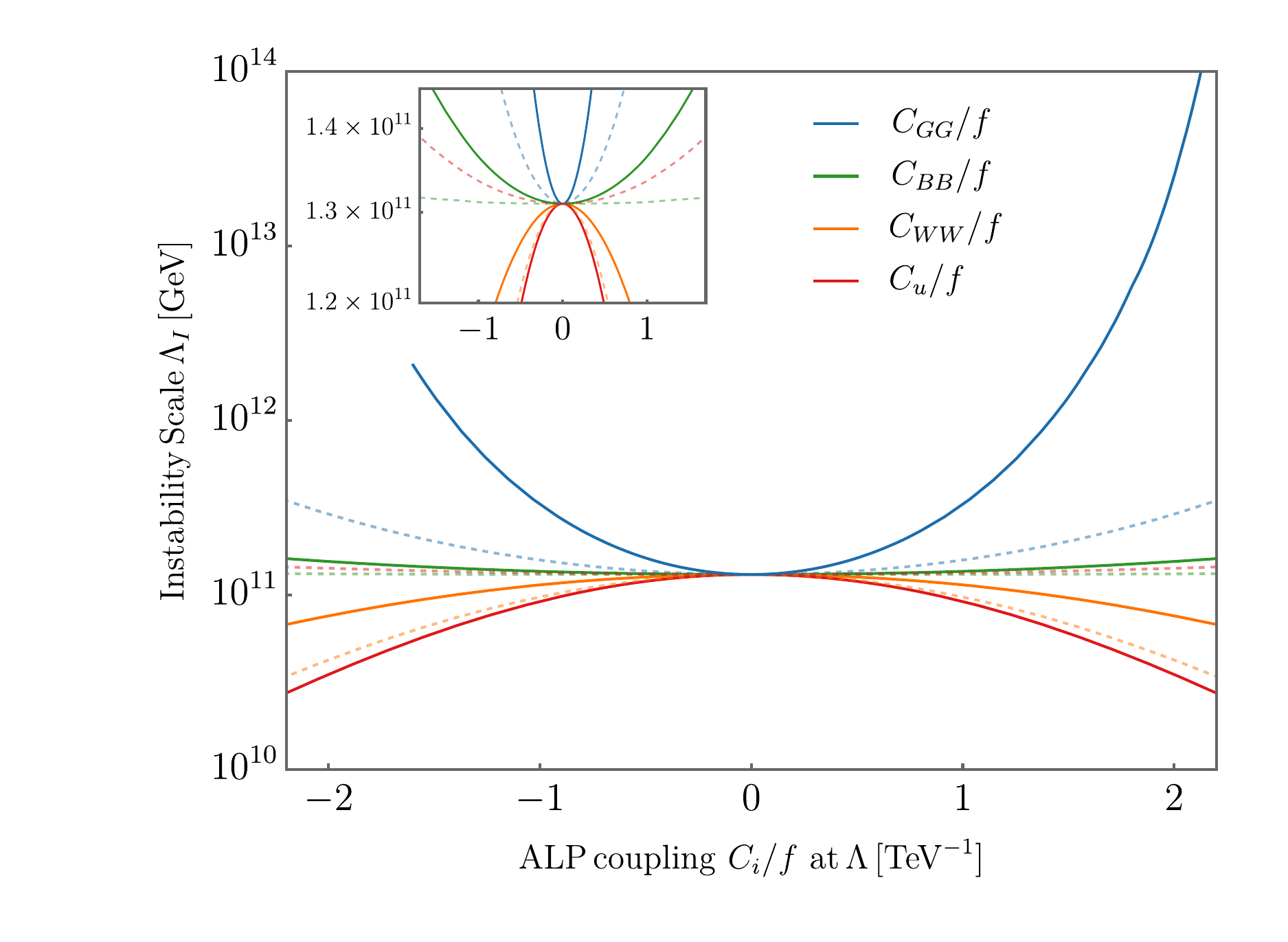}
\caption{Instability scale of the electroweak vacuum in the presence of nonzero ALP-SM couplings. The solid lines show the result for $m_a = 100$ GeV, while the dashed lines assume a vanishing ALP mass.}
\label{fig:Instability_Scale}
\end{figure}
\begin{itemize}
\setlength\itemsep{-0.1em}
    \item $\bm{C_{GG}}$: Any nonzero value of $C_{GG}$ will generally shift the instability scale towards higher values. While this coupling does not enter the RGE of $\lambda$ directly, its presence does not only generate nonzero SMEFT Wilson coefficients, but in particular also modifies the $\beta$-function of the strong coupling with an ALP-mass dependent term. In detail, 
    \begin{align}
        \frac{d \alpha_s}{d \ln\mu} \supset -16\,\alpha_s\frac{m_a^2}{\Lambda^2}\,C_{GG}^2\,.
    \end{align}
    As the dominant contribution from the strong coupling to the RG evolution of the Higgs quartic coupling is given by the term $ d \lambda/d \ln\mu \supset - \alpha_s\, [\bm{Y}_u]_{3, 3} $, the signs cancel, yielding an overall positive contribution to the RG evolution and thus shifting the instability scales to higher values. When the ALP mass is set to zero, the generation of SMEFT Wilson coefficients (in particular $C_{Hq}^{(3)}$), still pushes $\Lambda_I$ towards higher values, but the effect is much smaller than for $m_a = 100$ GeV, as can be seen in Figure~\ref{fig:Instability_Scale}. 
    \item $\bm{C_{WW}}$: In contrast to the ALP-gluon coupling described above, nonzero ALP couplings to $W$-bosons lead to a further destabilization of the electroweak vacuum. This is not surprising, since the square of this Wilson coefficient directly enters the RG evolution of $\lambda$ with a negative sign, see \eqref{eq:CWW_in_lambda}.
    \item $\bm{C_{u}}$: The coupling of ALPs to up-type quarks via $\bm{\tilde Y}_u =  i \,C_u \,\bm{Y}_u$ and $\bm{ Y^\prime}_u = C_u^2 \,\bm{Y}_u$ in the flavor-universal scenario, where $\bm{Y}_u$ is the up-type SM Yukawa matrix, has the largest destabilization effects among all ALP coefficients for $m_a = 100$ GeV, while it has nearly no impact for a vanishing ALP mass, as can be seen in the figure. The reason is that $C_u$ enters the RGE of $y_t$ via the term $\frac{d\,y_t}{d\ln\mu} \supset -\frac{m_a^2}{2\Lambda^2}\,C_u^2\,y_t $ in the flavor universal scenario, yielding a destabilizing effect.
    \item $\bm{C_{BB}}$: Similarly to the ALP-gluon case, the coupling $C_{BB}$ enters the scale evolution of $\lambda$ mainly via the SM gauge coupling $\alpha_1$ with 
        \begin{align}
        \frac{d \alpha_1}{d \ln\mu} \supset -16\,\alpha_1\frac{m_a^2}{\Lambda^2}\,C_{BB}^2\,.
    \end{align}
    The effect is, however, small in comparison to $C_{GG}$.
    \item $\bm{C_{e}},\,\bm{C_{d}},\,$: These ALP couplings have no significant effect on the scale evolution of $\lambda$.
\end{itemize}
\vspace{-.8cm}
\section{Bounds on the ALP couplings from the Higgs instability}
\label{sec:Instability_Regions}
\begin{figure}[h]
    \centering
 \includegraphics[width=.8\textwidth]{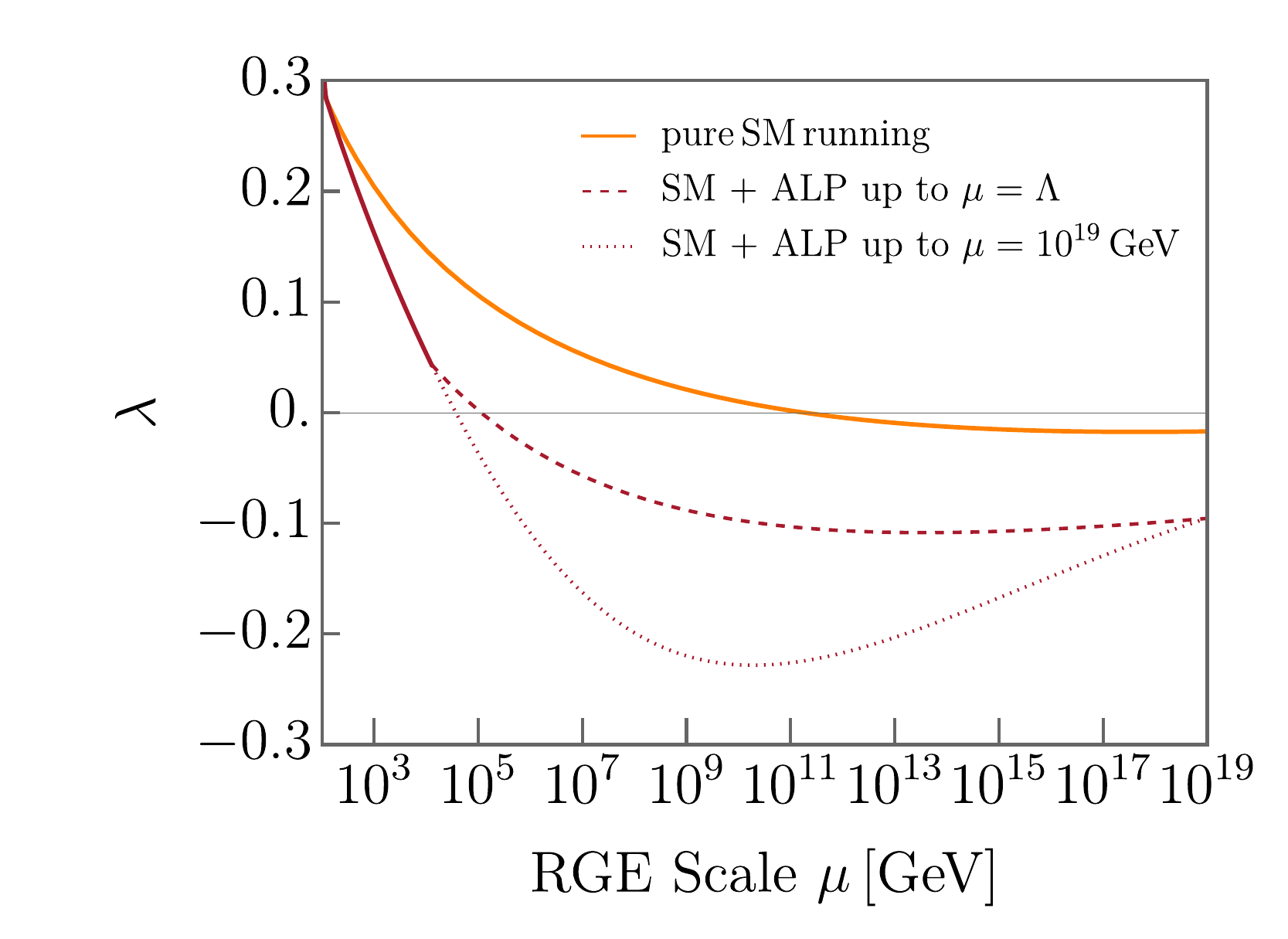}
\caption{Scale evolution of $\lambda$ for the case $C_{WW}/f = 12\,\text{TeV}^{-1}$, $m_a = 20\,$GeV. The red solid line shows the ALP + SM running until $\Lambda = 4\pi\,$TeV. Above this scale, the dashed red line is obtained by taking only SM effects into account above this scale, while the brown dotted line employs ALP effects on the running up to the Planck scale. The orange line shows the pure SM case (no ALP) for comparison.}
\label{fig:Running_to_Scale}
\end{figure}

\begin{figure}[H]
\centering
\includegraphics[width=.47\textwidth]{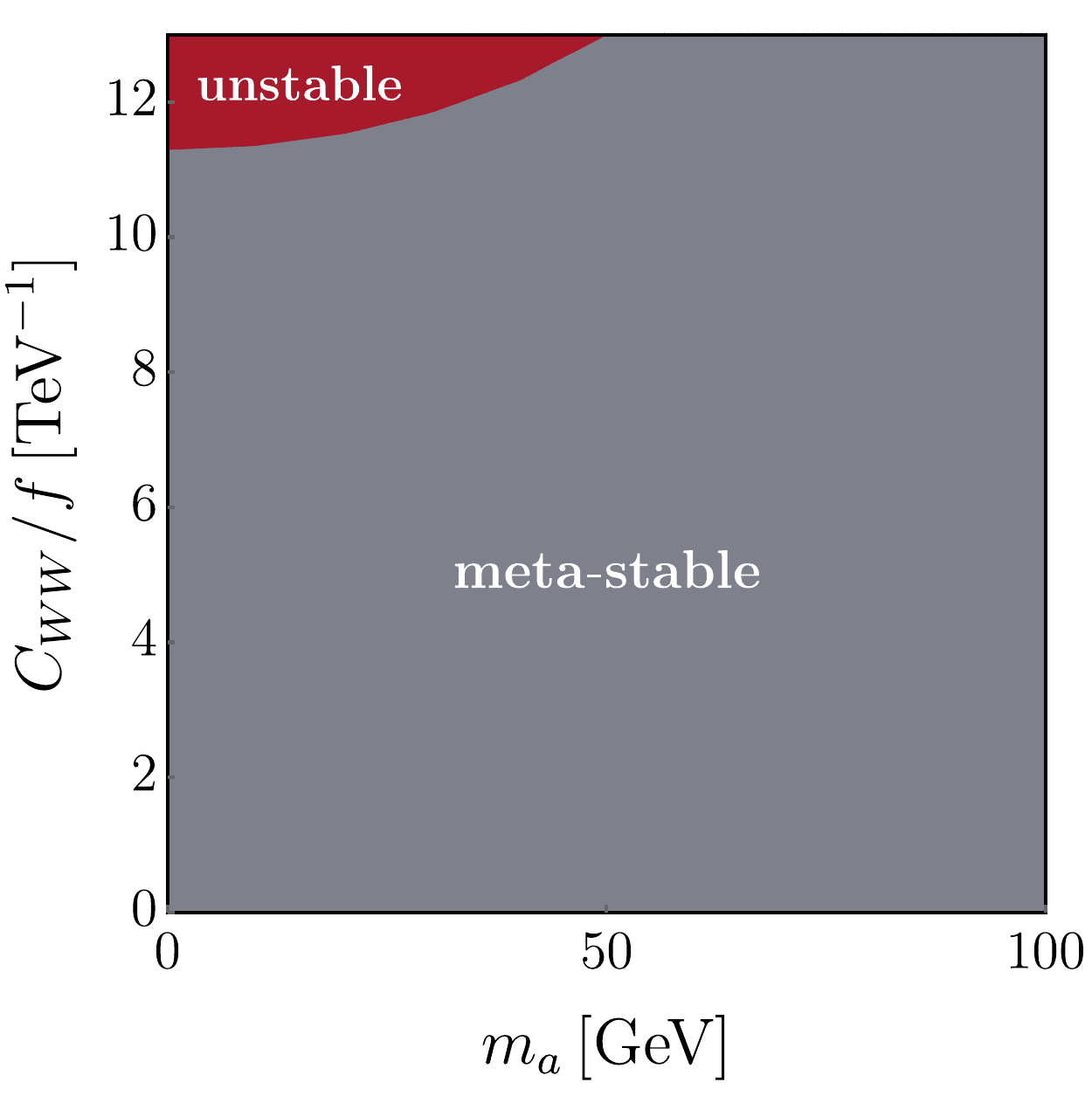} 
\includegraphics[width=.49\textwidth]{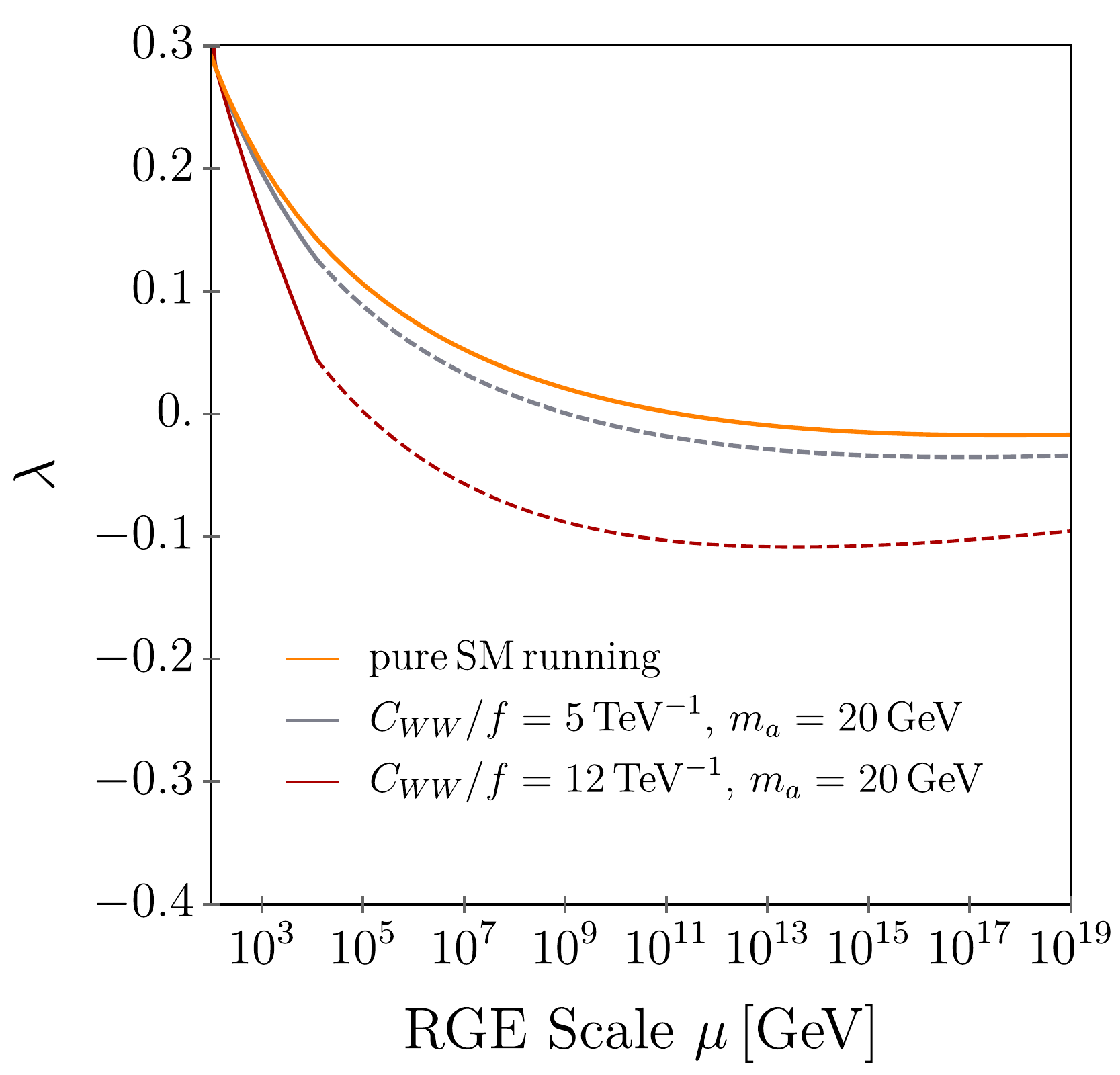}\\
\includegraphics[width=.47\textwidth]{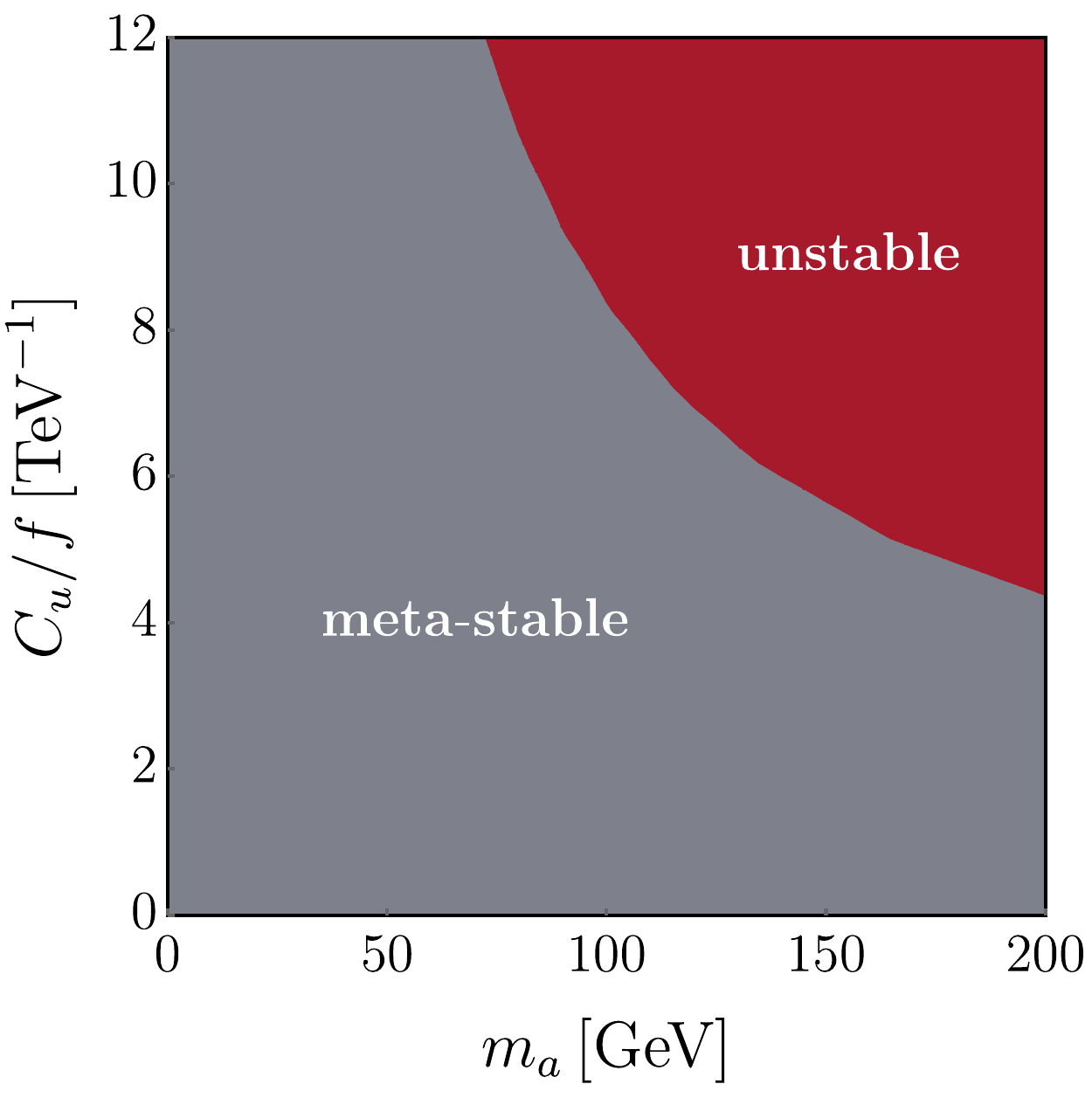}
\includegraphics[width=.49\textwidth]{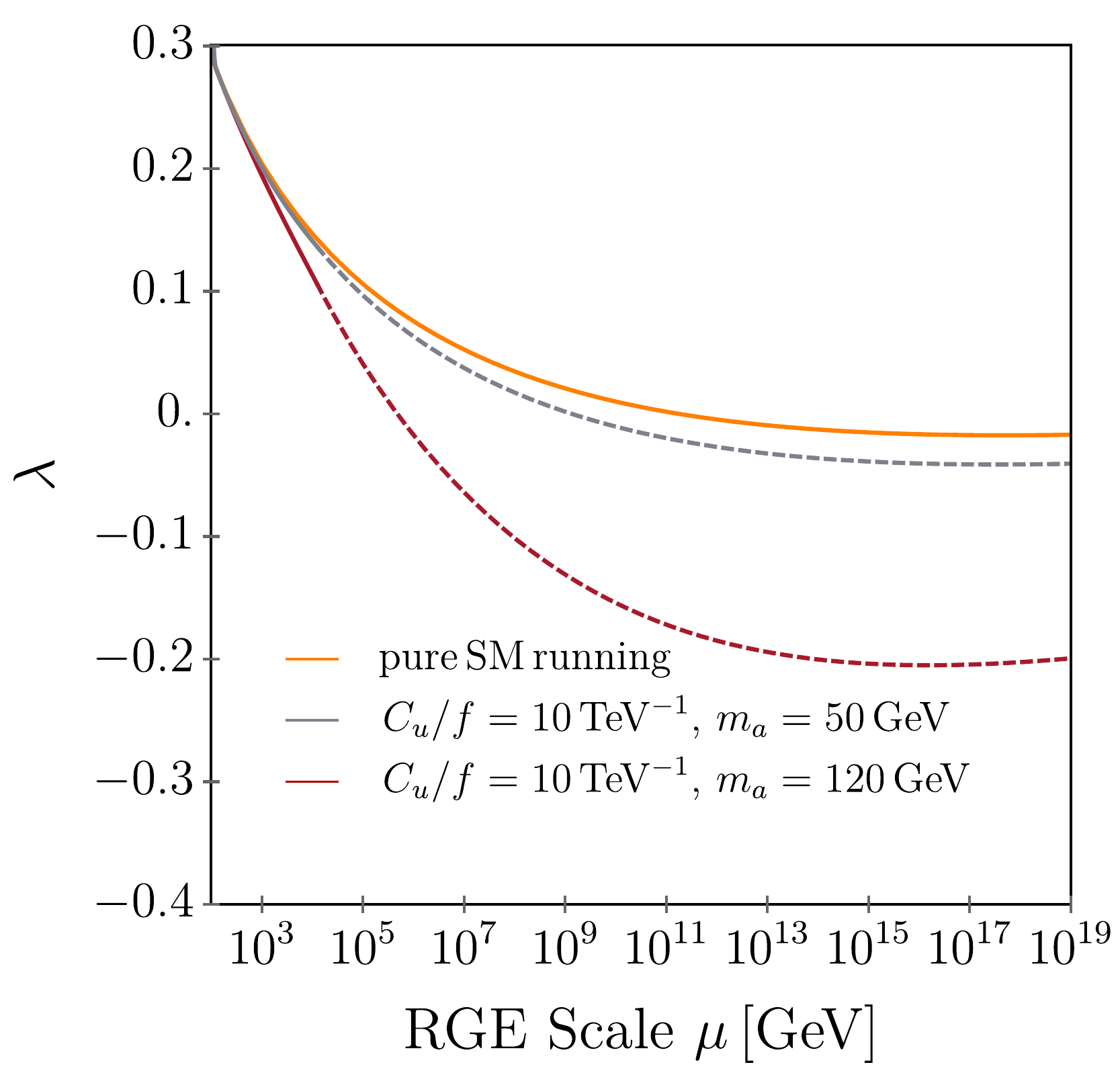}
\\
\includegraphics[width=.47\textwidth]{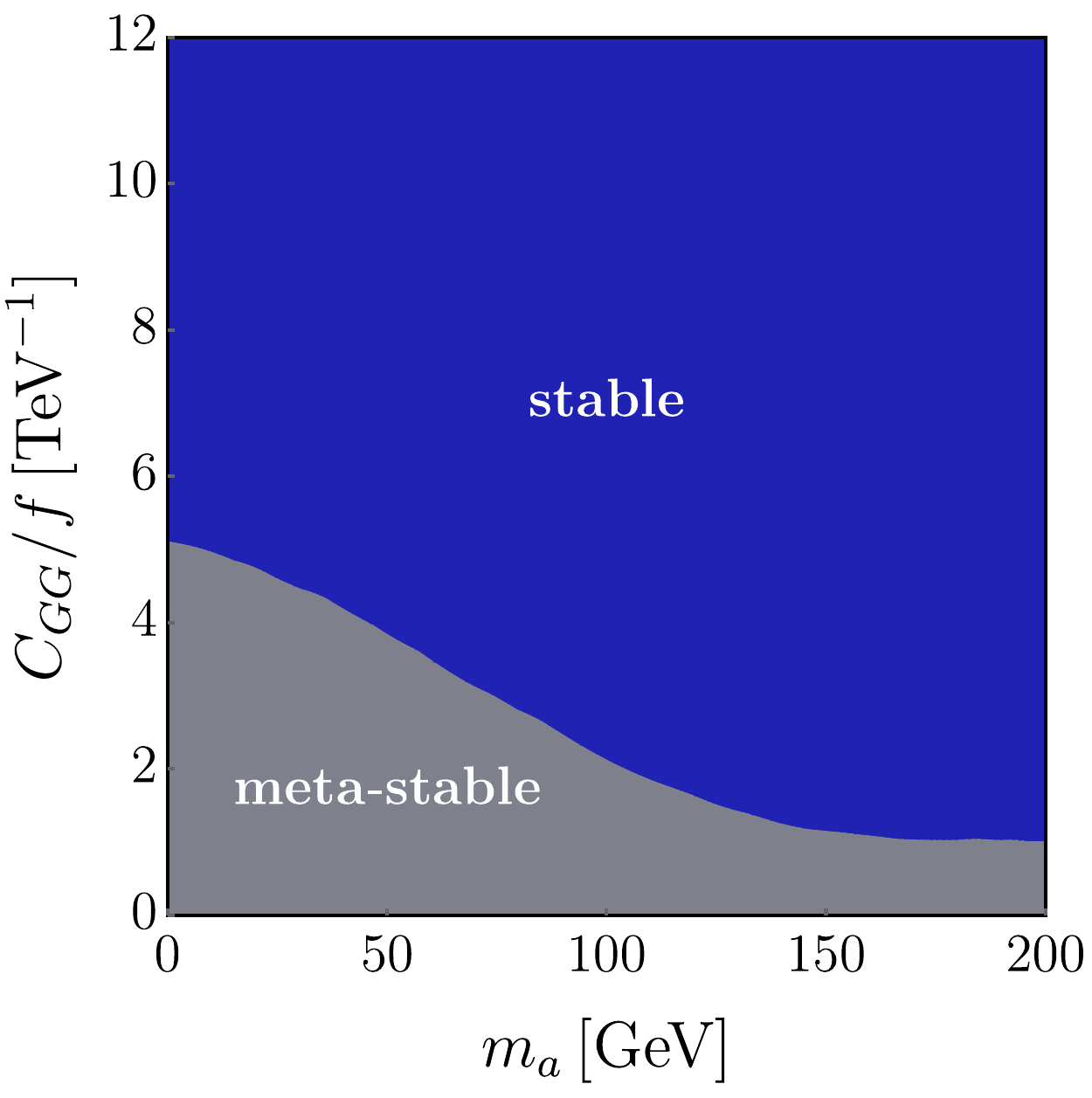}
\includegraphics[width=.49\textwidth]{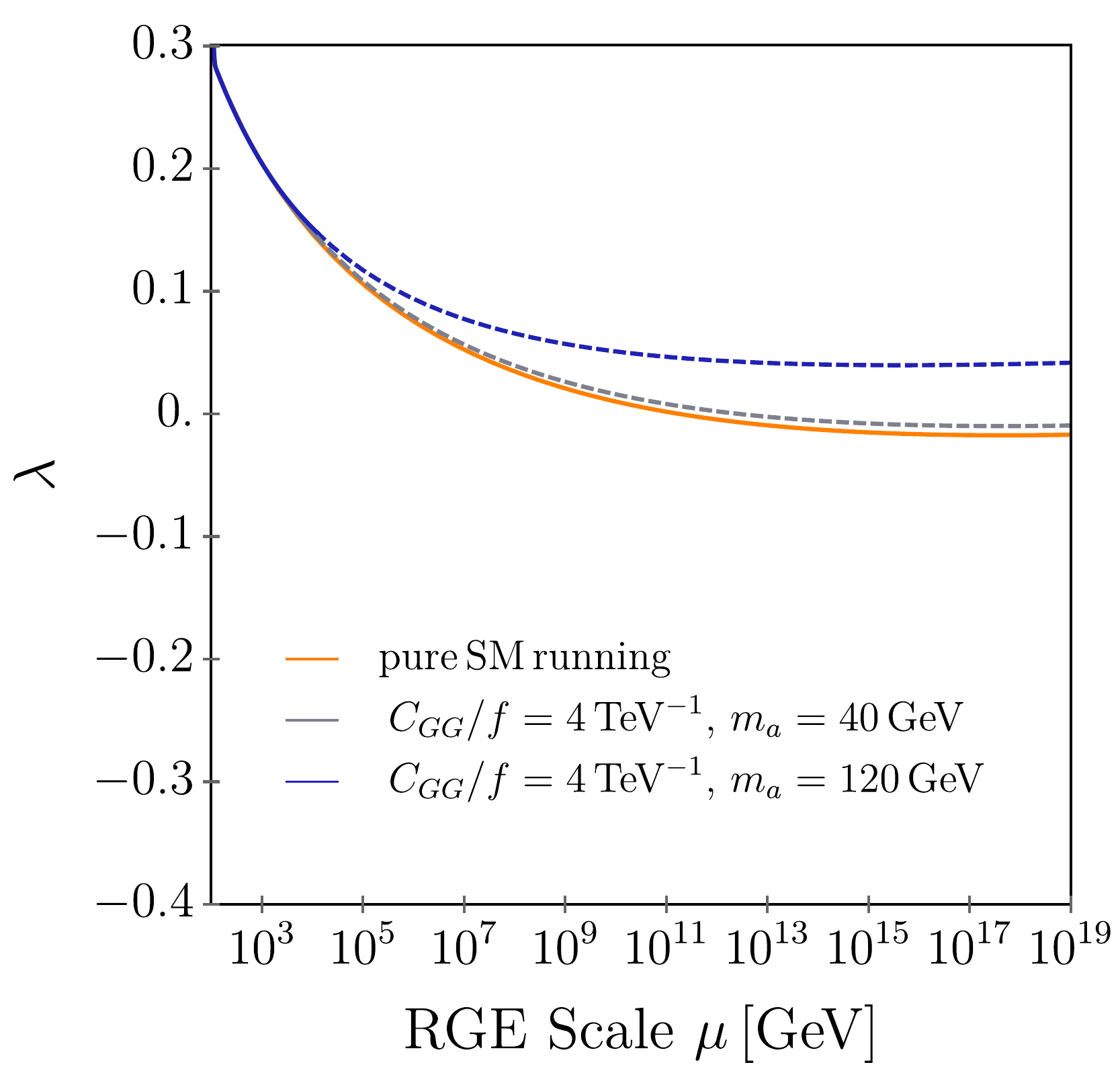}
\caption{Left: Unstable (red), stable (blue) and meta-stable (gray) regions for various ALP masses between $0$~GeV and $200$~GeV and one coupling turned on at the UV-scale, taken to  be $\Lambda = 4 \pi$~TeV. Right: Exemplary RG evolution of $\lambda$. The kink at $4 \pi $~TeV is a result from the generation of the ALP at this scale. The solid line shows the evolution below $4\pi$~TeV, where ALP effects modify the running.}
\label{fig:Ci_vs_ma}
\end{figure}

\begin{figure}[H]
\centering
\includegraphics[width=.45\textwidth]{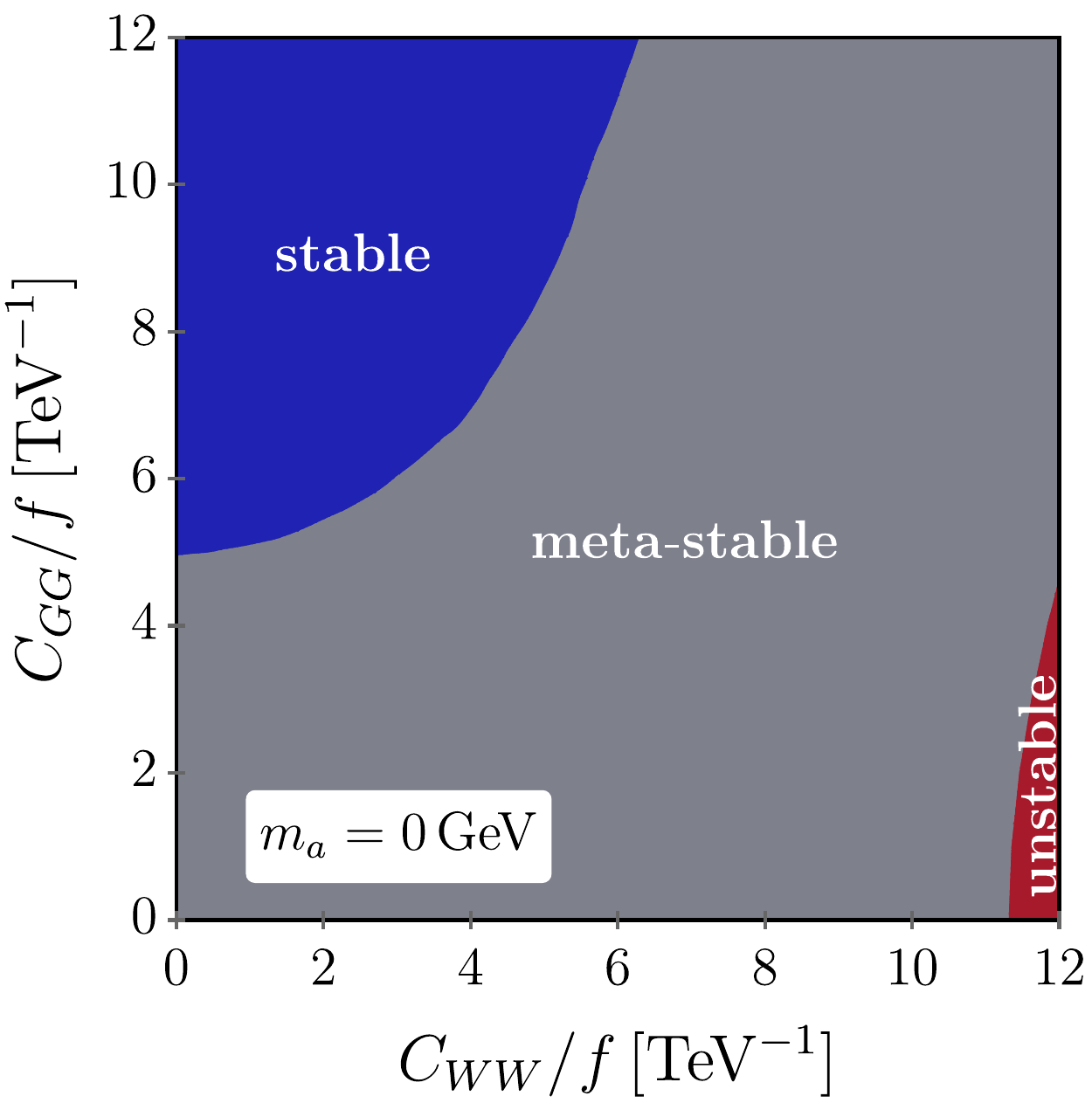}
\includegraphics[width=.45\textwidth]{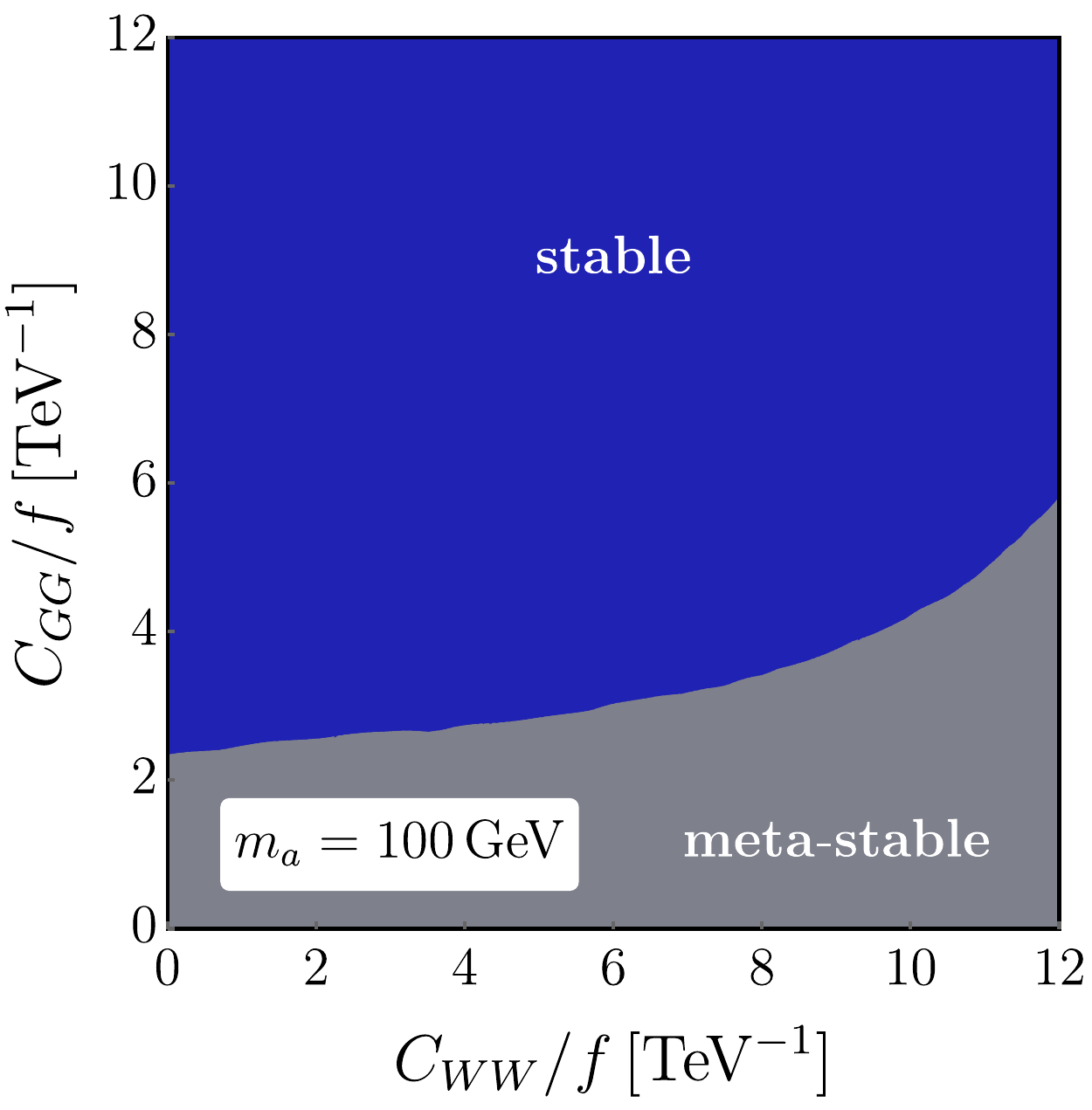}\\
\includegraphics[width=.45\textwidth]{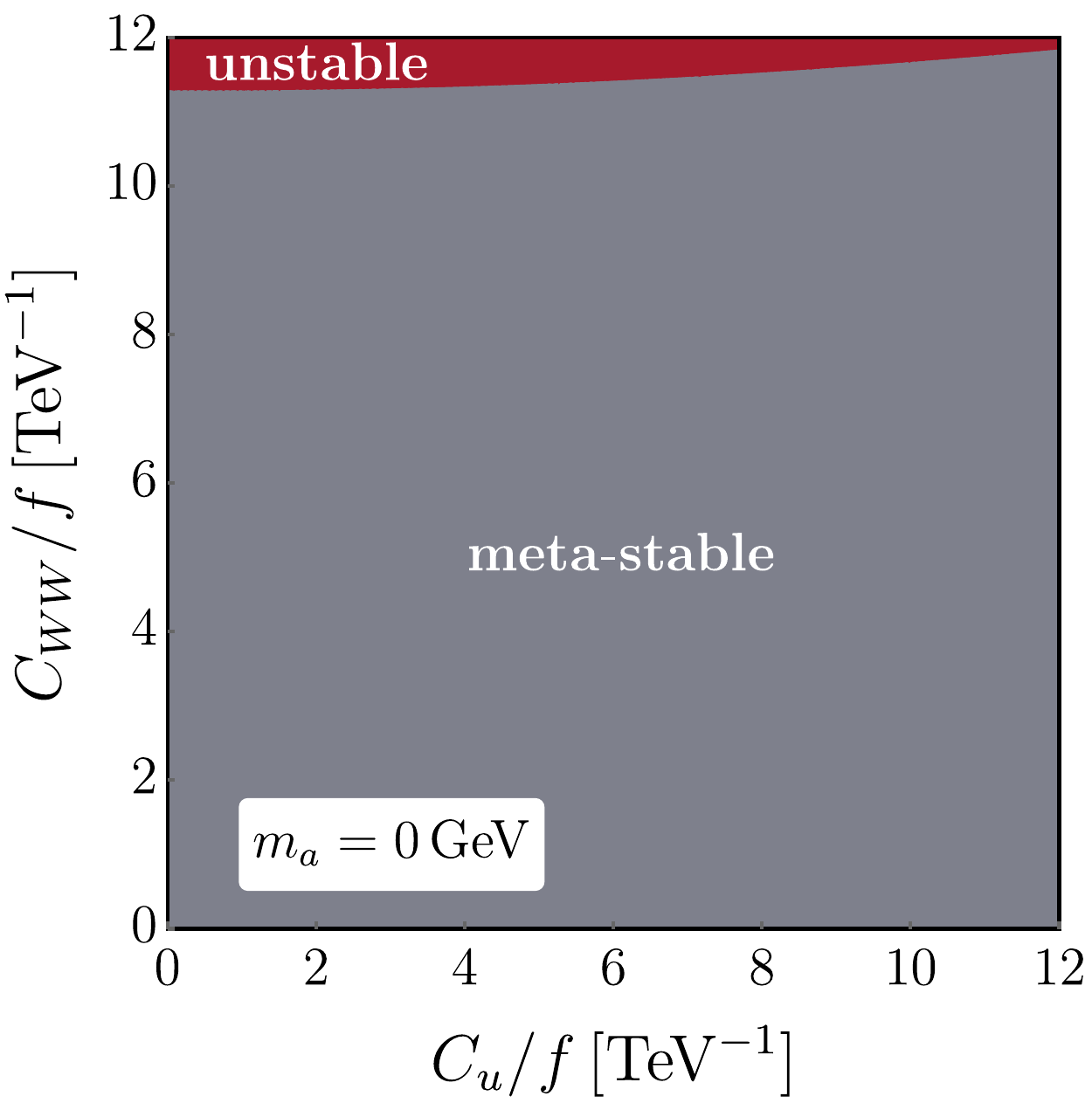}
\includegraphics[width=.45\textwidth]{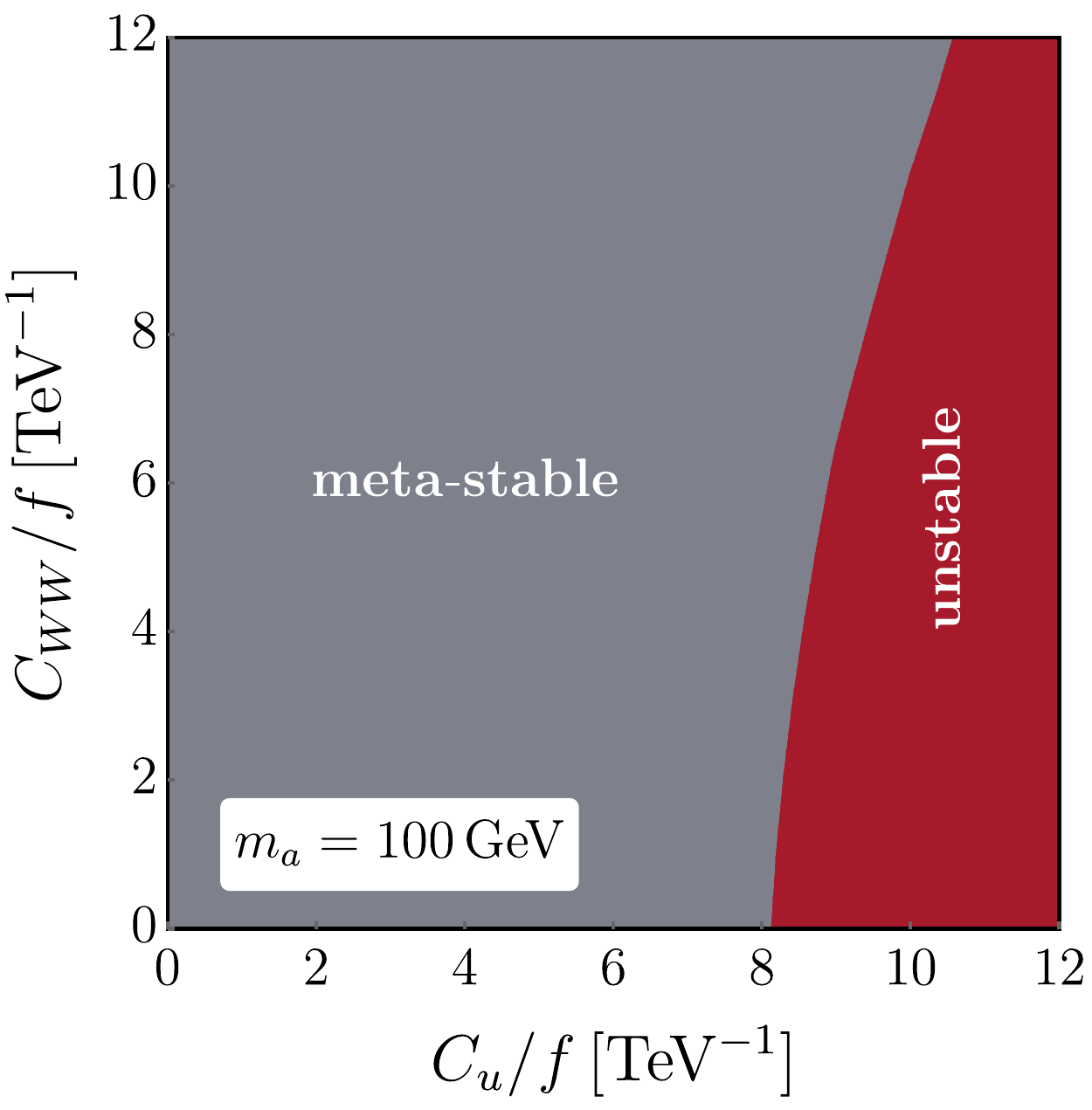}\\
\includegraphics[width=.45\textwidth]{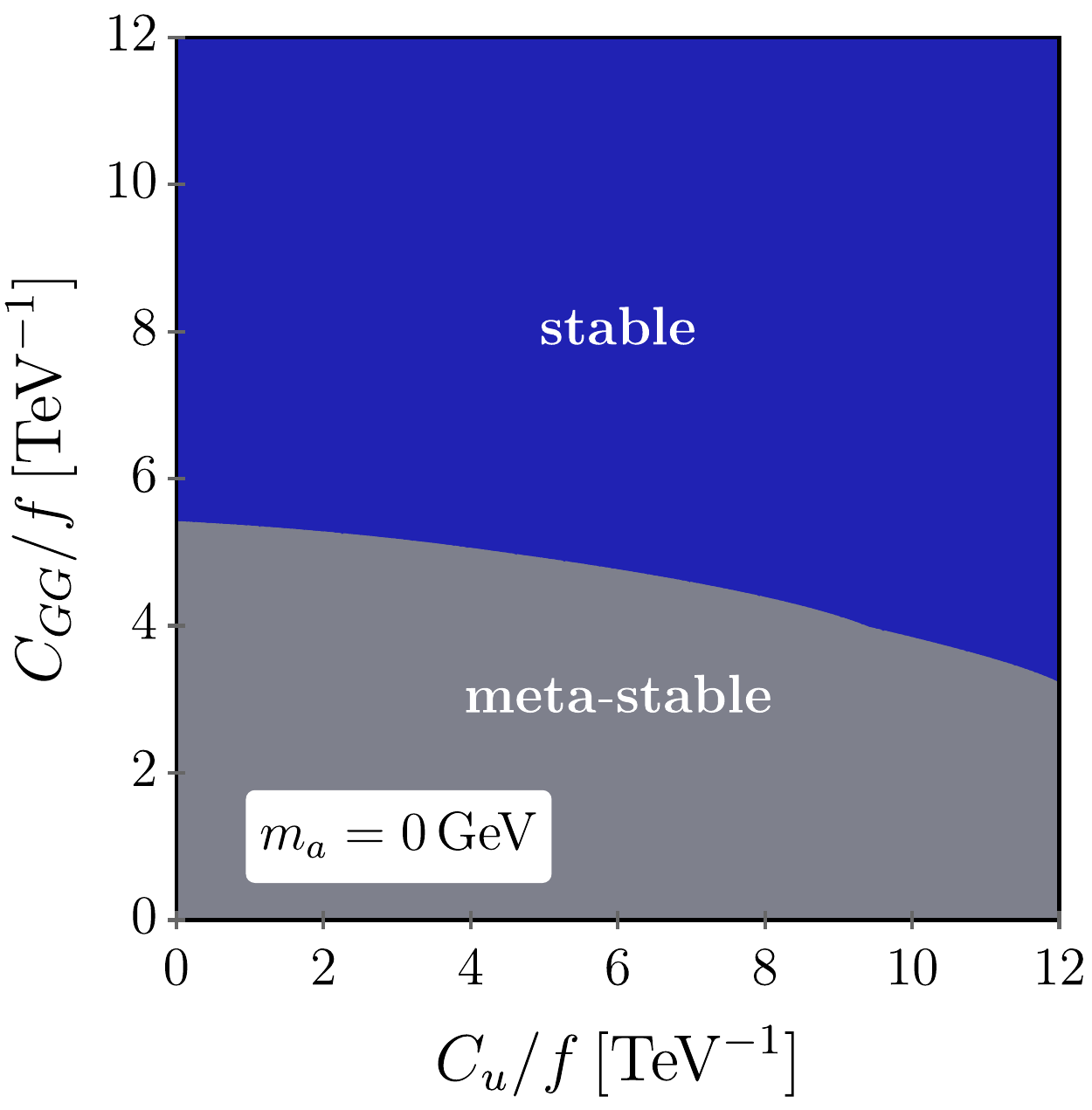}
\includegraphics[width=.45\textwidth]{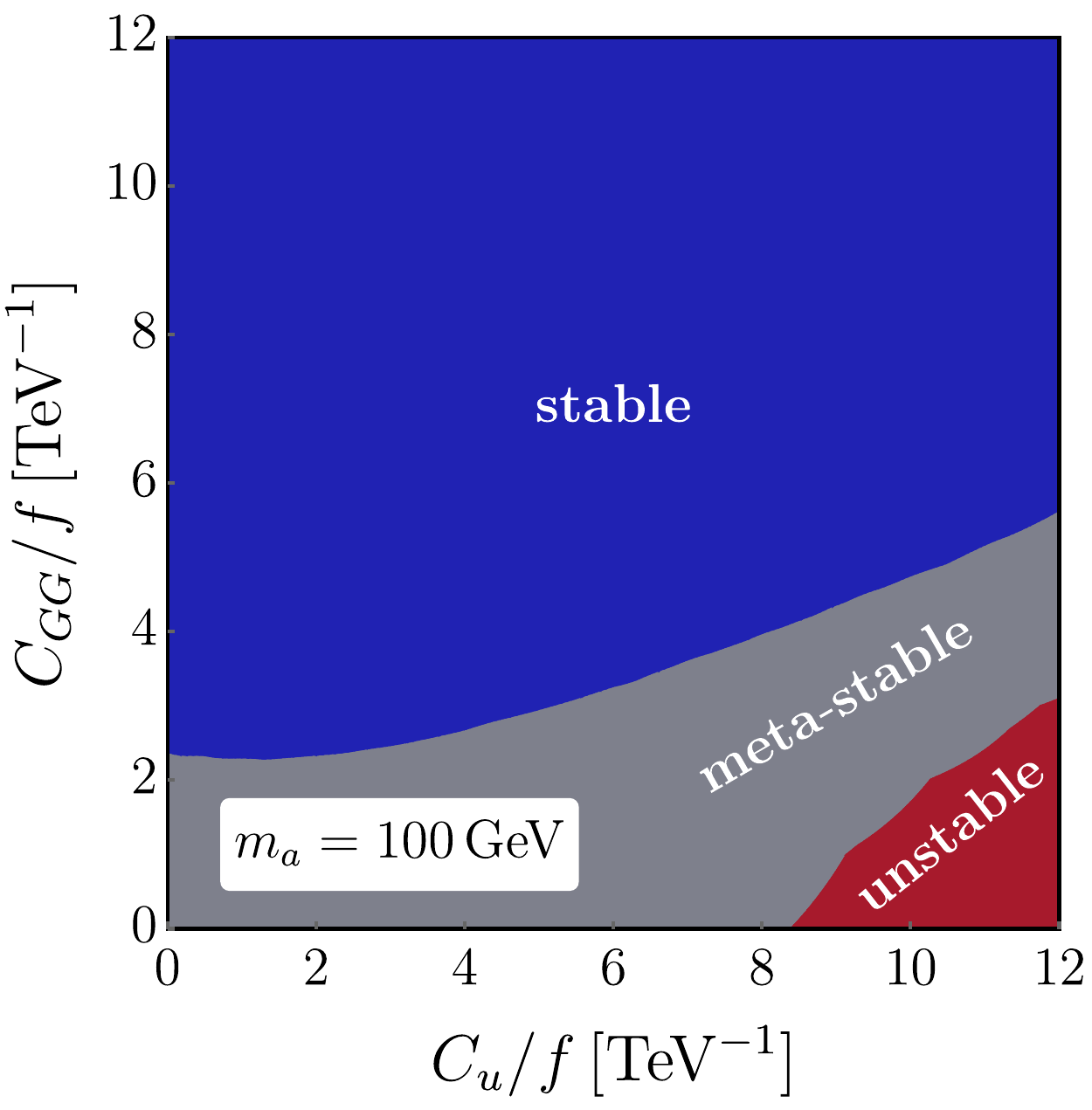}
\caption{Unstable (red), meta-stable (gray) and stable (blue) regions for various pairs of ALP couplings turned on at the scale $\Lambda = 4\pi$ TeV and a massless ALP mass. }
\label{fig:Ci_vs_Ci}
\end{figure}

We now proceed by scanning the ALP--SM couplings $C_i/f$ in the range $0 - 12\,\mathrm{TeV}^{-1}$ for $\Lambda = 4\pi$~TeV in order to identify those that lower the lifetime of the false electroweak vacuum to values below the age of the Universe. In general, effects of additional, model dependent degrees of freedom from heavy new physics can further modify the behavior of $\lambda$ above $\Lambda$, such that precise statements can only be obtained up to this scale. In order to consider the most conservative scenario, we use the ALP~+~SM RG evolution up to the scale of the $U(1)_A$ symmetry breaking, while ALP effects are not taken into account above $\Lambda$. A direct comparison of this scenario with the case in which the ALP effects are employed up to the Planck scale are depicted in Figure~\ref{fig:Running_to_Scale}. As can be inferred from the plot, the effect of the extended ALP running puts even tighter constraints on the couplings, as the instability becomes even more severe. \\
In Figure~\ref{fig:Ci_vs_ma} we show the results, obtained in this minimal scenario, for one nonzero coupling at the UV scale $\Lambda$ and the ALP mass $m_a$, which we vary between $0$ and $200$ GeV. For each plot, we also show the quartic Higgs coupling $\lambda$ as a function of the RG scale $\mu$ for one parameter point in the meta-stable and unstable region and the pure SM running without ALP as a reference. While the couplings $C_{WW}$ and $C_u$ have a destabilizing effect on the Higgs quartic coupling, as already seen in Section~\ref{sec:Instability_Scale}, high values of ALP--gluon couplings $C_{GG}$ at $\Lambda$ can even yield a stable Universe.\\ In Figure~\ref{fig:Ci_vs_Ci} we fix the ALP mass to either $0$ or $100$ GeV and show the regions of stability, meta-stability and instability for two nonzero coefficients at $\Lambda =  4 \pi $~TeV. Certainly, larger values of ALP couplings have stronger stabilizing or destabilizing effects on the electroweak vacuum. Correspondingly, the strongest effects are tightly constrained by experiment. While for low ALP masses, direct collider, flavor and astrophysical bounds provide the strongest constraints, the indirect bounds obtained in \cite{Biekotter:2023mpd} dominate for $m_a$ in the GeV range. It follows that coefficients required for the (de)stabilizing effects are experimentally excluded by several orders of magnitude for $m_a = 0$. For $m_a = 100\,\text{GeV}$, bounds for $C_{GG}/f$, $C_{WW}/f$, $C_{WW}/f$ and $C_{u}/f$ in $\text{TeV}^{-1}$ are of $\mathcal{O}(1)$. While the precise non-metastable regions presented in this paper lie in excluded regions, only small modifications from a UV completion of the ALP model are therefore sufficient to avoid these bounds.

\section{ALP unifying gauge forces?}
\label{sec:GUT}
\begin{figure}[H]
\centering
\includegraphics[width=.8\textwidth]{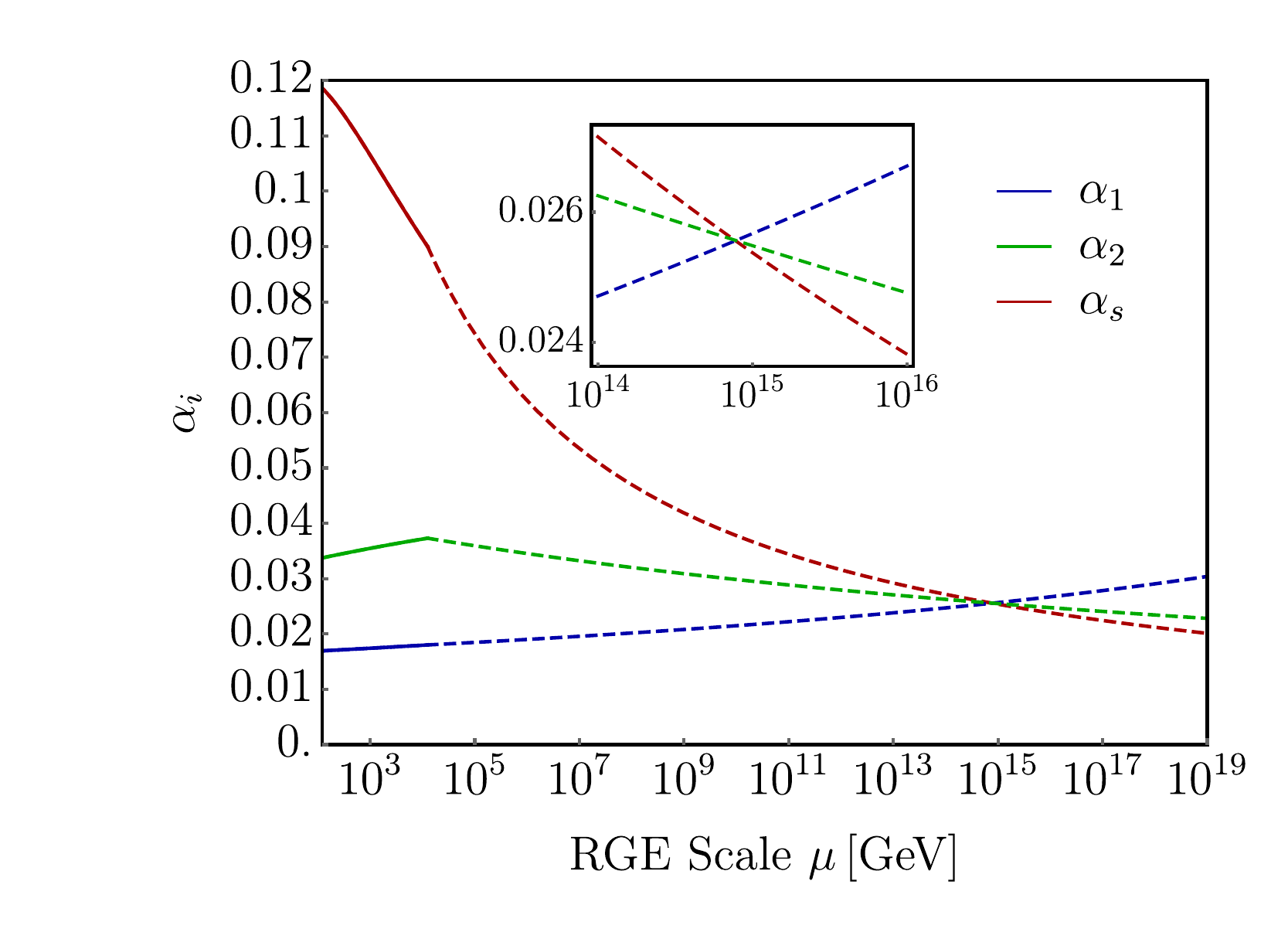}
\caption{RG evolution of $\alpha_1,\,\alpha_2$ and $\alpha_3$ in the presence of an ALP with $m_a = 120$ GeV, $\Lambda = 4\pi$ TeV and $C_{GG} = 5.2\,\mathrm{TeV^{-1}},\,C_{WW} = 5.0\,\mathrm{TeV^{-1}},\, C_{BB} =1.0\,\mathrm{TeV^{-1}}$. The solid line shows the evolution below $4\pi$~TeV, where ALP effects modify the running, while the dashed lines above $\Lambda$ would in principle be subject to additional new physics in concrete ALP models. }
\label{fig:GUT}
\end{figure}

In addition to the Higgs quartic coupling $\lambda$, ALP--SM interactions also modify the $\beta$-functions of the three gauge couplings. Defining $ d\alpha_s/d\ln\mu\equiv-2\alpha_s\,\beta^{(3)}(\{\alpha_i\})$ and equivalently for $\alpha_1$ and $\alpha_2$, ALP interactions modify the SM contributions directly via \cite{Galda:2021hbr}
\begin{align}
    \beta^{(i)}(\{\alpha_i\}) &= \beta_{\mathrm{SM}}^{(1)}(\{\alpha_i\})  + \frac{40}{3}\frac{ m_a^2}{\Lambda^2}\,C_{BB}^2\,,\notag\\
    \beta^{(2)}(\{\alpha_i\}) &= \beta_{\mathrm{SM}}^{(2)}(\{\alpha_i\})  + \frac{8 m_a^2}{\Lambda^2}\,C_{WW}^2\,,\notag\\
    \beta^{(3)}(\{\alpha_i\}) &= \beta_{\mathrm{SM}}^{(3)}(\{\alpha_i\})  + \frac{8 m_a^2}{\Lambda^2}\,C_{GG}^2\,,
\end{align}
where in the first line the $SU(5)$ GUT normalization, $g_1 = \sqrt{5/3}\,g_Y$, was used \cite{Buttazzo:2013uya}.
Any ALP with nonzero mass will thus necessarily have an impact on the RG evolution of the gauge couplings. An example for a specific ALP-SM configuration at the scale of global symmetry breaking $\Lambda = 4 \pi$ TeV that results in a unification of $\alpha_1,\,\alpha_2$ and $\alpha_3$ at around $10^{15}$ GeV is shown in Figure~\ref{fig:GUT}. Here, we choose $m_a = 120$~GeV, as well as the ALP-boson couplings at $\Lambda$ as $C_{GG} = 5.2\,\mathrm{TeV^{-1}},\,C_{WW} = 5.0\,\mathrm{TeV^{-1}},\, C_{BB} =1.0\,\mathrm{TeV^{-1}}$.

\section{Conclusions}
While the initial motivation for axions was to find an elegant solution for the strong CP problem of the SM, these particles, along with the more general axion-like particles, have become well-motivated candidates to solve a variety of fundamental questions, ranging from dark matter to the flavor and hierarchy problems. In this paper we have shown that by virtue of the modification of the $\beta$-functions of the SM parameters, in particular the RG equations of the Higgs quartic coupling $\lambda$, the gauge coupling $g_s$ and the top-Yukawa coupling $y_t$, the list of potential solutions can be extended by two: first, the ALP--gluon coupling has the potential of shifting $\lambda$ to positive values, thereby avoiding that the electroweak vacuum is a metastable state that eventually decays to a lower, true ground state.
On the other hand, we have shown that several ALP couplings, in particular, $C_{WW}$ and $C_u$, tend to drive the Higgs quartic coupling to values lower than in the SM. This effect can be used to constrain these parameters, under the assumption that the electroweak vacuum's lifetime must exceed the age of the Universe.\\
Second, the modifications of the $\beta$-functions of the three SM gauge couplings from $C_{GG},\,C_{WW}, \,C_{BB}$ and the ALP mass can, in certain configurations, lead to a unification of those values at high energies around $\mu \approx 10^{15}$ GeV. \\
In this paper we remained agnostic about the underlying UV model that generates the ALP. In such concrete scenarios, threshold corrections and additional modifications to the scale evolution of $\lambda$ can arise, which would also need to be taken into account. However, we showed that even in the conservative approach presented here, where we focus on the effects from only an ALP with weak scale couplings, significant changes in the Higgs vacuum stability and the gauge couplings can be obtained.

\subsection*{Acknowledgments}
We thank Matthias König for cross-checking the RGE of $C_{HH}$ with an unpublished version of \texttt{Matchete} \cite{Fuentes-Martin:2022jrf}. The research of A.M.G.\ and M.N.\ was supported by the Cluster of Excellence \textit{Precision Physics, Fundamental Interactions, and Structure of Matter} (PRISMA$^+$, EXC 2118/1) within the German Excellence Strategy (Project-ID 390831469). The Feynman diagrams in this paper have been drawn with the LaTeX package \texttt{TikZ-Feynman} \cite{Ellis:2016jkw}.

\newpage
\bibliographystyle{JHEP}
\bibliography{references}

\providecommand{\href}[2]{#2}\begingroup\raggedright\begin{thebibliography}{10}

\bibitem{Peccei:1977hh}
R.~D. Peccei and H.~R. Quinn, \emph{{CP Conservation in the Presence of
  Instantons}}, \href{https://doi.org/10.1103/PhysRevLett.38.1440}{\emph{Phys.
  Rev. Lett.} {\bfseries 38} (1977) 1440--1443}.

\bibitem{Weinberg:1977ma}
S.~Weinberg, \emph{{A New Light Boson?}},
  \href{https://doi.org/10.1103/PhysRevLett.40.223}{\emph{Phys. Rev. Lett.}
  {\bfseries 40} (1978) 223--226}.

\bibitem{Wilczek:1977pj}
F.~Wilczek, \emph{{Problem of Strong $P$ and $T$ Invariance in the Presence of
  Instantons}}, \href{https://doi.org/10.1103/PhysRevLett.40.279}{\emph{Phys.
  Rev. Lett.} {\bfseries 40} (1978) 279--282}.

\bibitem{Calibbi:2016hwq}
L.~Calibbi, F.~Goertz, D.~Redigolo, R.~Ziegler and J.~Zupan, \emph{{Minimal
  axion model from flavor}},
  \href{https://doi.org/10.1103/PhysRevD.95.095009}{\emph{Phys. Rev. D}
  {\bfseries 95} (2017) 095009},
  [\href{https://arxiv.org/abs/1612.08040}{{\ttfamily 1612.08040}}].

\bibitem{Ema:2016ops}
Y.~Ema, K.~Hamaguchi, T.~Moroi and K.~Nakayama, \emph{{Flaxion: a minimal
  extension to solve puzzles in the standard model}},
  \href{https://doi.org/10.1007/JHEP01(2017)096}{\emph{JHEP} {\bfseries 01}
  (2017) 096}, [\href{https://arxiv.org/abs/1612.05492}{{\ttfamily
  1612.05492}}].

\bibitem{Bagger:1994hh}
J.~Bagger, E.~Poppitz and L.~Randall, \emph{{The R axion from dynamical
  supersymmetry breaking}},
  \href{https://doi.org/10.1016/0550-3213(94)90123-6}{\emph{Nucl. Phys. B}
  {\bfseries 426} (1994) 3--18},
  [\href{https://arxiv.org/abs/hep-ph/9405345}{{\ttfamily hep-ph/9405345}}].

\bibitem{Gripaios:2009pe}
B.~Gripaios, A.~Pomarol, F.~Riva and J.~Serra, \emph{{Beyond the Minimal
  Composite Higgs Model}},
  \href{https://doi.org/10.1088/1126-6708/2009/04/070}{\emph{JHEP} {\bfseries
  04} (2009) 070}, [\href{https://arxiv.org/abs/0902.1483}{{\ttfamily
  0902.1483}}].

\bibitem{Ferretti:2013kya}
G.~Ferretti and D.~Karateev, \emph{{Fermionic UV completions of Composite Higgs
  models}}, \href{https://doi.org/10.1007/JHEP03(2014)077}{\emph{JHEP}
  {\bfseries 03} (2014) 077},
  [\href{https://arxiv.org/abs/1312.5330}{{\ttfamily 1312.5330}}].

\bibitem{Graham:2015cka}
P.~W. Graham, D.~E. Kaplan and S.~Rajendran, \emph{{Cosmological Relaxation of
  the Electroweak Scale}},
  \href{https://doi.org/10.1103/PhysRevLett.115.221801}{\emph{Phys. Rev. Lett.}
  {\bfseries 115} (2015) 221801},
  [\href{https://arxiv.org/abs/1504.07551}{{\ttfamily 1504.07551}}].

\bibitem{Bellazzini:2017neg}
B.~Bellazzini, A.~Mariotti, D.~Redigolo, F.~Sala and J.~Serra, \emph{{$R$-axion
  at colliders}},
  \href{https://doi.org/10.1103/PhysRevLett.119.141804}{\emph{Phys. Rev. Lett.}
  {\bfseries 119} (2017) 141804},
  [\href{https://arxiv.org/abs/1702.02152}{{\ttfamily 1702.02152}}].

\bibitem{Galda:2021hbr}
A.~M. Galda, M.~Neubert and S.~Renner, \emph{{ALP \textemdash{} SMEFT
  interference}}, \href{https://doi.org/10.1007/JHEP06(2021)135}{\emph{JHEP}
  {\bfseries 06} (2021) 135},
  [\href{https://arxiv.org/abs/2105.01078}{{\ttfamily 2105.01078}}].

\bibitem{Degrassi:2012ry}
G.~Degrassi, S.~Di~Vita, J.~Elias-Miro, J.~R. Espinosa, G.~F. Giudice,
  G.~Isidori et~al., \emph{{Higgs mass and vacuum stability in the Standard
  Model at NNLO}}, \href{https://doi.org/10.1007/JHEP08(2012)098}{\emph{JHEP}
  {\bfseries 08} (2012) 098},
  [\href{https://arxiv.org/abs/1205.6497}{{\ttfamily 1205.6497}}].

\bibitem{Georgi:1986df}
H.~Georgi, D.~B. Kaplan and L.~Randall, \emph{{Manifesting the Invisible Axion
  at Low-energies}},
  \href{https://doi.org/10.1016/0370-2693(86)90688-X}{\emph{Phys. Lett. B}
  {\bfseries 169} (1986) 73--78}.

\bibitem{Bauer:2020jbp}
M.~Bauer, M.~Neubert, S.~Renner, M.~Schnubel and A.~Thamm, \emph{{The
  Low-Energy Effective Theory of Axions and ALPs}},
  \href{https://doi.org/10.1007/JHEP04(2021)063}{\emph{JHEP} {\bfseries 04}
  (2021) 063}, [\href{https://arxiv.org/abs/2012.12272}{{\ttfamily
  2012.12272}}].

\bibitem{Biekotter:2023mpd}
A.~Biek\"otter, J.~Fuentes-Mart\'\i{}n, A.~M. Galda and M.~Neubert, \emph{{A
  global analysis of axion-like particle interactions using SMEFT fits}},
  \href{https://doi.org/10.1007/JHEP09(2023)120}{\emph{JHEP} {\bfseries 09}
  (2023) 120}, [\href{https://arxiv.org/abs/2307.10372}{{\ttfamily
  2307.10372}}].

\bibitem{Jenkins:2013wua}
E.~E. Jenkins, A.~V. Manohar and M.~Trott, \emph{{Renormalization Group
  Evolution of the Standard Model Dimension Six Operators II: Yukawa
  Dependence}}, \href{https://doi.org/10.1007/JHEP01(2014)035}{\emph{JHEP}
  {\bfseries 01} (2014) 035},
  [\href{https://arxiv.org/abs/1310.4838}{{\ttfamily 1310.4838}}].

\bibitem{ParticleDataGroup:2022pth}
{\scshape Particle Data Group} collaboration, R.~L. Workman et~al.,
  \emph{{Review of Particle Physics}},
  \href{https://doi.org/10.1093/ptep/ptac097}{\emph{PTEP} {\bfseries 2022}
  (2022) 083C01}.

\bibitem{Jenkins:2013zja}
E.~E. Jenkins, A.~V. Manohar and M.~Trott, \emph{{Renormalization Group
  Evolution of the Standard Model Dimension Six Operators I: Formalism and
  lambda Dependence}},
  \href{https://doi.org/10.1007/JHEP10(2013)087}{\emph{JHEP} {\bfseries 10}
  (2013) 087}, [\href{https://arxiv.org/abs/1308.2627}{{\ttfamily 1308.2627}}].

\bibitem{Alonso:2013hga}
R.~Alonso, E.~E. Jenkins, A.~V. Manohar and M.~Trott, \emph{{Renormalization
  Group Evolution of the Standard Model Dimension Six Operators III: Gauge
  Coupling Dependence and Phenomenology}},
  \href{https://doi.org/10.1007/JHEP04(2014)159}{\emph{JHEP} {\bfseries 04}
  (2014) 159}, [\href{https://arxiv.org/abs/1312.2014}{{\ttfamily 1312.2014}}].

\bibitem{Buttazzo:2013uya}
D.~Buttazzo, G.~Degrassi, P.~P. Giardino, G.~F. Giudice, F.~Sala, A.~Salvio
  et~al., \emph{{Investigating the near-criticality of the Higgs boson}},
  \href{https://doi.org/10.1007/JHEP12(2013)089}{\emph{JHEP} {\bfseries 12}
  (2013) 089}, [\href{https://arxiv.org/abs/1307.3536}{{\ttfamily 1307.3536}}].

\bibitem{Celis:2017hod}
A.~Celis, J.~Fuentes-Martin, A.~Vicente and J.~Virto, \emph{{DsixTools: The
  Standard Model Effective Field Theory Toolkit}},
  \href{https://doi.org/10.1140/epjc/s10052-017-4967-6}{\emph{Eur. Phys. J. C}
  {\bfseries 77} (2017) 405},
  [\href{https://arxiv.org/abs/1704.04504}{{\ttfamily 1704.04504}}].

\bibitem{Fuentes-Martin:2020zaz}
J.~Fuentes-Martin, P.~Ruiz-Femenia, A.~Vicente and J.~Virto, \emph{{DsixTools
  2.0: The Effective Field Theory Toolkit}},
  \href{https://doi.org/10.1140/epjc/s10052-020-08778-y}{\emph{Eur. Phys. J. C}
  {\bfseries 81} (2021) 167},
  [\href{https://arxiv.org/abs/2010.16341}{{\ttfamily 2010.16341}}].

\bibitem{Fuentes-Martin:2022jrf}
J.~Fuentes-Mart\'\i{}n, M.~K\"onig, J.~Pag\`es, A.~E. Thomsen and F.~Wilsch,
  \emph{{A proof of concept for matchete: an automated tool for matching
  effective theories}},
  \href{https://doi.org/10.1140/epjc/s10052-023-11726-1}{\emph{Eur. Phys. J. C}
  {\bfseries 83} (2023) 662},
  [\href{https://arxiv.org/abs/2212.04510}{{\ttfamily 2212.04510}}].

\bibitem{Ellis:2016jkw}
J.~Ellis, \emph{{TikZ-Feynman: Feynman diagrams with TikZ}},
  \href{https://doi.org/10.1016/j.cpc.2016.08.019}{\emph{Comput. Phys. Commun.}
  {\bfseries 210} (2017) 103--123},
  [\href{https://arxiv.org/abs/1601.05437}{{\ttfamily 1601.05437}}].

\end{thebibliography}\endgroup
\end{document}